\documentclass[twocolumn,showpacs]{revtex4}

\usepackage{bm}

\usepackage{graphicx}
\usepackage{subfigure}

\newcommand{\ww}{\stackrel{\rm wrong}{\rm way}}
\newcommand{\wrongp}{\stackrel{\rm wrong}{\rm phase}}

\begin{document}

\title{Anomalies of Wave-Particle Duality due
to Translational-Internal Entanglement}

\author{Michal Kol\'{a}\v{r}}

\affiliation{Department of Theoretical Physics, Palack\'{y}
University, 17. listopadu 50,
77200 Olomouc, Czech Republic}

\author{Tom\'{a}\v{s} Opatrn\'{y}}

\affiliation{Department of Theoretical Physics, Palack\'{y}
University, 17. listopadu 50,
77200 Olomouc, Czech Republic}

\author{Nir Bar-Gill and Gershon Kurizki}

\affiliation{Weizmann Institute of Science, 76100
Rehovot, Israel}

\date{\today }

\pacs{03.65.Ud, 03.65.Vf, 03.75.Dg}

\begin{abstract}
We predict that if internal and momentum states of an
interfering object are correlated (entangled), then by measuring its
internal state we may {\em infer} both path (corpuscular) and phase
(wavelike) information with {\em much higher precision} than for objects
lacking such entanglement. We thereby partly circumvent the standard complementarity
constraints of which-path detection.
\end{abstract}

\maketitle

Our ability to know the actual path taken by a diffracting or interfering
particle has been debated since the early days of quantum mechanics: suffice it
to recall the Bohr--Einstein ``which-path'' controversy in the context of
two--slit interference \cite{Zurek}. More recent analysis of this issue has
prompted the formulation \cite{Wooters,ScullyNature} and subsequent testing
\cite{RempeNature,Mei} of the fundamental complementarity relation $D^2 + V^2 \le 1$
between path distinguishability $D$, which is the certitude of knowing the
actual path of an interfering particle  by coupling a detector to one of these
paths, and the fringe visibility (contrast) $V$, which determines the ability to
infer the phase difference between the alternative paths. But must
``which-path'' information always come at the expense of interference-phase
information? We show that this is {\em not necessarily} the case when internal
and momentum states of the interfering particle become {\em entangled}. Then, by
measuring its internal state we may infer both path (corpuscular) and phase
(wavelike) information with much higher precision than for objects lacking such
entanglement, thereby partly {\em circumventing} the complementarity constraints. This
anomaly may yield novel interferometric applications.

Consider spin-1/2 particles (or their analogs: two-level atoms) of
mass $M$ that are prepared in the four--dimensional (4D) input
state
\begin{equation}
|\psi _{\mathrm{input}}\rangle =\left( \sqrt{1-p} |k_1
\rangle |1\rangle + \sqrt{p} |k_2 \rangle |2\rangle \right). \label{e1}
\end{equation}
Here $\sqrt{1-p}$ and $\sqrt{p}$ are the probability amplitudes (chosen to be real and positive) of the internal states $| 1 \rangle,| 2 \rangle$,
which correspond to the
internal energy levels $\epsilon_1,\epsilon_2$. These states are assumed to have $x$-oriented momenta, $\hbar k_1,\hbar k_2$,
constrained by the total energy of
state (\ref{e1}):
\begin{equation}
E = \frac{\hbar^2 k_1^2}{2 M} + \epsilon_1 =
\frac{\hbar^2 k_2^2}{2 M} + \epsilon_2. \label{e1b}
\end{equation}
The choice (\ref{e1b}) enforces a {\it stationary}
(time-independent) {\em scenario} in what follows. We use
Eq.(\ref{e1}) as a short-hand description of narrow-momentum
wavepackets (gaussians) whose coherence length along $x$ exceeds the
size of the interferometric setup $L$. This implies that
$|k_{1(2)},1(2) \rangle$ really represent $\int dk f_{1(2)} (k) |k,
1(2) \rangle$, with gaussian distributions $f_{1(2)} (k) \propto
\exp(-|k-k_{1(2)}|^2/\Delta_x^2)$, such that $\Delta_x \ll 1/L$.

We are interested in the peculiar spatial properties of the {\it stationary}
state (\ref{e1}), which exhibits a feature that has hitherto not been studied in
the context of single-particle interferometry: Bell-like \cite{Scully} {\em quantum
correlation} between two momentum states and two nondegenerate internal states,
hereafter named translational-internal entanglement (TIE). This entanglement vanishes for $p=0,1$ and is maximal for $p=1/2$. The realization of TIE is
discussed later on.

\begin{figure}[tbh]
\centering \hspace{-0.06\linewidth}
\includegraphics[width=.9\linewidth]{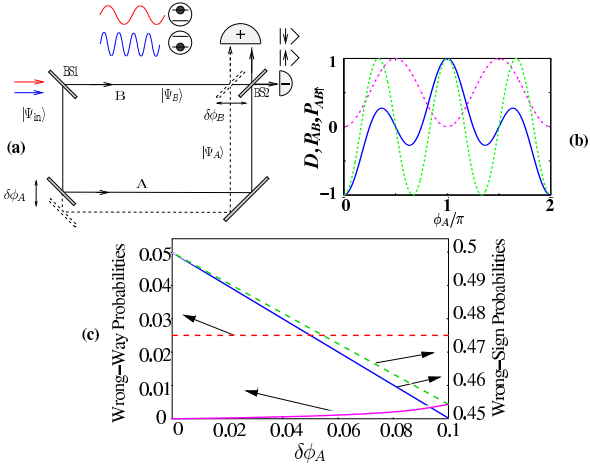}
\caption{(a) A particle in the TIE state (\ref{e1}) in a MZI. It
traverses the interferometer from BS1 to BS2 via paths A and B,
whose mean phases, $\phi_{A1(2)}=k_{1(2)} L_A$ and
$\phi_{B1(2)}=k_{1(2)} L_B$, fluctuate by $(\delta \phi_A)_{1(2)}$
and $(\delta \phi_B)_{1(2)}$, respectively.  Both output detectors
$+$ and $-$ discriminate internal states $|\uparrow\rangle$ and $|
\downarrow \rangle$ (eq. (\ref{e6})).
(b) The $\phi_{A1}$ dependence of TIE functions. (eqs. (\ref{e7}),
(\ref{e10})) for $p=1/2, k_2=3 k_1$, $\phi_{B1}=\pi$: non-sinusoidal
$P_{AB\uparrow}$ (solid-bold-blue), $D_{\rm TIE}$
(dash-dot,magenta); sinusoidal $P_{AB}$ (dashed,green, $V=1$) for $3
k_1=3 k_2=k_{\rm max}$ (standard case). (c) Wrong-way and
wrong-phase probabilities for TIE (solid magenta and solid blue,
respectively - Eq. (\ref{e10})) are lower than for the standard case (dashed
red and dashed green, respectively - Eq. (\ref{e11a}) for
$D_S=0.95$) as functions of $| \delta \phi_A |$. } \label{f1}
\end{figure}

We will show that the TIE state (\ref{e1}) yields much more information than unentangled states on propagation along {\em both arms} of the
simple Mach-Zehnder interferometer \cite{Scully} (MZI). The wavefunction
$\langle x| \psi_{\mathrm{input}} \rangle = \left( \sqrt{1-p} e^{i k_1
x} |1 \rangle + \sqrt{p} e^{i k_2 x} | 2 \rangle \right)$ is ``split'' at the balanced
(50\%-50\%) input beam splitter (BS1) into two beams that propagate along either
of the two arms of length $L_A$ or $L_B$, then recombine at the 50\%-50\% beam
merger BS2 (Fig.~\ref{f1}a). Propagating these beams along the two arms, we find
that right before the beam merger the ``final'' wavefunction is, in the $x$
representation
\begin{eqnarray}
| \psi_f \rangle_x &=& \frac{1}{\sqrt{2}}
\left( | \psi_{Af} \rangle_x |A \rangle +
| \psi_{Bf} \rangle_x |B \rangle \right), \label{e2} \\
\nonumber | \psi_{Af} \rangle_x &\equiv& \langle x | \psi_{Af}
\rangle = \left( \sqrt{1-p} e^{i \phi_{A1}}
|1 \rangle_A + \sqrt{p} e^{i \phi_{A2}} |2 \rangle_A \right), \nonumber \\
\nonumber | \psi_{Bf} \rangle_x &\equiv& \langle x | \psi_{Bf}
\rangle = \left( \sqrt{1-p} e^{i \phi_{B1}} |1 \rangle_B + \sqrt{p}
e^{i \phi_{B2}} |2 \rangle_B \right).\\ \label{e3}
\end{eqnarray}
In (\ref{e2}), we have introduced $|A \rangle$ and $|B \rangle$, the {\it
spatially--orthogonal states} representing the respective paths (which means
that the spatial width of the wave packet perpendicular to the propagation ($x$-)
axis is much smaller than the distance between arms A and B). We have also
introduced in (\ref{e3}) the phases $\phi_{A1(2)}=k_{1(2)} L_{A}$,
$\phi_{B1(2)}=k_{1(2)} L_{B}$. As $k_{1(2)}L_A,k_{1(2)}L_B \rightarrow 2m \pi$, we recover from
(\ref{e3}) the input states in arms A,B. We see that a single-arm contribution
to the wavefunction, $| \psi_{Af} \rangle$ or $| \psi_{Bf} \rangle$, is rotated
by the phase $\phi_{A2}-\phi_{A1}=(k_2-k_1) L_A$ or
$\phi_{B2}-\phi_{B1}=(k_2-k_1) L_B$, respectively. These phases, representing the interference of $|k_1 \rangle |1 \rangle$ and
$|k_2 \rangle |2 \rangle$, distinguish TIE from standard states: they are {\it
``which-path'' markers travelling with the particle}, encoding the path traversed along each arm in the superposition of internal
states $|1\rangle$ and $|2\rangle$, as in a Ramsey interferometer \cite{Scully}.

This feature of TIE allows us to record the interference of {\em
four} distinct contributions of paths A and B (rather than the usual
two). It will be shown that by reading out this interference at the
output detectors, we may infer, with much higher certitude than for
standard states, parameters whose knowledge is usually {\em
complementary}: the path (arm) state ($|A \rangle$ or $|B \rangle$)
{\em and} the phase (length) difference of the two interfering arms.
At the output detectors $+$ and $-$ (Fig.~\ref{f1}a), the
wavefunctions are, respectively, the sum and the difference of the
$A$ and $B$ path amplitudes \cite{Scully}
\begin{eqnarray}
| \psi_{\pm} \rangle &=& \frac{1}{2} \left( | \psi_{Af}
\rangle_x \pm | \psi_{Bf} \rangle_x \right)|\pm\rangle \nonumber \\
&=& \frac{1}{2} \left[ \sqrt{1-p} \left( e^{i \phi_{A1}}
\pm e^{i \phi_{B1}} \right) |1 \rangle \nonumber \right. \\
&+& \left. \sqrt{p} \left( e^{i \phi_{A2}} \pm e^{i \phi_{B2}} \right)
|2 \rangle \right]|\pm\rangle. \label{e4}
\end{eqnarray}

We henceforth choose, for the sake of concreteness, $k_2 = 3 k_1$,
denoting $\phi_{A1}-\phi_{B1} \equiv \phi, \phi_{A2}-\phi_{B2} = 3
\phi$. If we now ignore the internal states of the particle, we get
the following click probabilities at the +/$-$ detectors,
\begin{equation}
P_{\pm} = \langle \psi_{\pm} | \psi_{\pm} \rangle \equiv \frac{1}{2}
\left[ 1 \pm (1-p) \cos \phi \pm p \cos 3 \phi \right]. \label{e5}
\end{equation}
This procedure yields no which-path (arm) information, only phase-difference information.

In order to gain both path and phase information, our output detectors +/$-$
should discriminate between internal states, such that each
detector projects onto one of the orthogonal states
\begin{eqnarray}
| \uparrow \rangle &=& \frac{1}{\sqrt{2}}
\left( | 1 \rangle +  | 2 \rangle \right), \nonumber \\
| \downarrow \rangle &=& \frac{1}{\sqrt{2}}
\left(  | 1 \rangle - | 2 \rangle \right). \label{e6}
\end{eqnarray}
This setup has four different output (detection) channels: $+ \uparrow, +
\downarrow, - \uparrow, - \downarrow$.
The $+ \uparrow$ and $- \uparrow$ channel
probabilities are obtained upon projecting the wavefunctions $| \psi_{\pm}
\rangle$ onto $| \uparrow \rangle$. We shall analyze them explicitly for maximum TIE, $p=1/2$, which will be shown to be the optimal choice. These probabilities are:
\begin{eqnarray}
P_{\pm \uparrow} & = & \frac{1}{2}
\left( P_{A \uparrow} + P_{B \uparrow}
\pm P_{AB \uparrow} \right), \nonumber \\
P_{A \uparrow} &=& \frac{1}{4} \left[ 1 + \cos \left( 2 \phi_A
 \right) \right], \nonumber \\
P_{B \uparrow} &=& \frac{1}{4} \left[ 1 + \cos \left( 2 \phi_B
 \right) \right], \nonumber \\
P_{AB \uparrow} &=& \frac{1}{4} \left[ \cos \phi + \cos 3 \phi
+ \cos \left( 3 \phi_B - \phi_A  \right) \right. \nonumber \\
&+& \left. \cos \left( 3 \phi_A - \phi_B  \right) \right].
\label{e7}
\end{eqnarray}
Here $P_{A \uparrow}$ denotes the joint probability of finding the particle
right before BS2 in arm A in the internal state $\uparrow$, and similarly for
$P_{B \uparrow}$. The $P_{AB \uparrow}$ term is the interference contribution to
the $\uparrow$ channel. The $\downarrow$ channel probabilities are obtained from
(\ref{e7}) upon replacing + by $-$  in front of all cosine
terms except for cos $\phi$ and cos $3 \phi$.

Equation (\ref{e7}) implies that, due to
TIE, the single-arm contributions $P_{A \uparrow}$ and $P_{B
\uparrow}$ (or their $\downarrow$ counterparts) have sinusoidal
dependence on the phases, as compared to the {\it non-sinusoidal},
complicated phase dependence of the arm-interference contribution
$P_{AB \uparrow}$ (see Fig.~1b). This difference in phase dependence
will be shown to be crucial for inferring the path (arm) together
with the small phase deviations (path-length deviations) $\pm
\delta\phi_{A(B)}$ around their mean values $\overline{\phi_{A(B)}}
\equiv k_1 \overline{L}_{A(B)}$.

If $N_{\rm in}$ particles travel through the MZI, then we obtain from (\ref{e7})
the ``imbalance'' $( \delta N)_{\uparrow}$ between the counts of the two output
detectors and the total number $(N_{\rm tot})_{\uparrow}$ counts at both
detectors in the $\uparrow$ channel. As an example we choose
$\overline{\phi_{A(B)}}$ such that $P_{B \uparrow} \approx 1/2$,  $P_{A
\uparrow} \approx 0$: $\overline\phi_{B}=\pi$,
$\overline\phi_{A}=\frac{\pi}{2}$, and consider small length deviations $\delta
\phi_A \equiv k_1 \delta L_A$, $\delta \phi_B \equiv k_1 \delta L_B$. Then
\begin{eqnarray}
\nonumber
(N_+)_{\uparrow} - (N_-)_{\uparrow} &\equiv&
\left( \delta N \right)_{\uparrow} = N_{\rm in}
\left( P_{+ \uparrow}-P_{- \uparrow} \right) \nonumber \\
\nonumber
&=&  N_{\rm in}P_{AB\uparrow}\approx
- N_{\rm in}\delta\phi_A , \label{e8} \\
\nonumber
(N_+)_{\uparrow} + (N_-)_{\uparrow} &\equiv&
\left( N_{\rm tot} \right)_{\uparrow} =
 N_{\rm in} (P_{A \uparrow} + P_{B \uparrow}) \nonumber \\
&\approx& \frac{N_{\rm in}}{2}
\left( 1 - \delta \phi_B^2 + \delta \phi_A^2 \right).
\label{e9a}
\end{eqnarray}
Hence, upon varying $\delta \phi_A = k_1 \delta L_A$, we may deduce that
$(N_{\rm tot})_{\uparrow}$ has a contribution from the unlikely path $N_{\rm in}
P_{A \uparrow}$, which scales {\em quadratically} with $\delta \phi_A$, whereas
$(\delta N)_{\uparrow}$, which is proportional to the path-interference
probability $(P_{AB})_{\uparrow}$ of $| \uparrow \rangle$ particles, depends
{\it linearly} on $\delta \phi_A$. The same conclusions apply to the
$\downarrow$ channels upon exchanging A and B, whereupon $P_{A \downarrow}
\approx 1/2 - \delta \phi_A^2/2$, $P_{B \downarrow} \approx \delta \phi_B^2 /
2$, $P_{AB\downarrow} \approx  \delta \phi_B$. Thus, we conclude that the
particle has most likely traversed path A if it is found in state $|\downarrow \rangle$
or path B if it is found in $|\uparrow \rangle$.

The information on the likelihood of traversing paths A, B or both,
embodied by Eqs.(\ref{e7}), (\ref{e8}), is inferred {\it without}
which-path detection inside the MZI. But is the concept of likely
(``correct'') or unlikely (``wrong'') paths meaningful at all here?
As we will detail elsewhere, this concept is meaningful, since we may
{\it verify} our inferences for small subensembles, and compare the
errors of these inferences to those obtainable in the standard case
by conventional which-path detection.
The probabilities of our wrong-way and wrong-phase guesses
per particle are expected from (\ref{e7}), (\ref{e8}) to be (see
Fig.~1c)
\begin{eqnarray}
\left( P_{\ww} \right)_{\rm TIE} &=&  P_{A \uparrow} +  P_{B \downarrow}\approx
\frac{\delta \phi_A^2}{2}, \label{e10} \\
\left( P_{\wrongp} \right)_{\rm TIE}
 &=& P_{+}(-|\delta \phi_A|)
 = P_{-}( +|\delta \phi_A|) \nonumber \\
&\approx&  \frac{1}{2} (1 - | \delta \phi_A| ).\nonumber
\end{eqnarray}

To extract both path and phase information in the absence of TIE,
when $k \equiv k_1=k_2$, we are forced to adopt the {\it standard}
recipe \cite{Zurek,Wooters,ScullyNature,RempeNature} of placing a
detector in one of the arms inside the MZI. Let this detector be
imperfect, allowing path distinguishability $D_S<1$. This path
distinguishability is phase-{\it independent}, i.e. it does not
depend on $\phi = k (L_A-L_B)$. The interference at the output
detector then oscillates as
\begin{eqnarray}
\left( P_\pm \right)_S=\frac{1}{2}\pm\frac{V \cos\phi}{2},
\label{pab}
\end{eqnarray}
the visibility $V$ being complementary to the distinguishability
\cite{ScullyNature}:
\begin{equation}
V^2 + D_S^2=1.
\label{equality}
\end{equation}
The counterparts of (\ref{e10}) are then the error probabilities
(see Fig.~1c):
\begin{eqnarray}
\left( P_{\ww} \right)_S &=& \frac{1-D_S}{2},\label{e11a}\\
\nonumber
\left( P_{\wrongp} \right)_S
&\approx & \frac{1}{2} (1 - \sqrt{1-D_S^2} |\delta \phi|).
\end{eqnarray}
It can be checked that the TIE-based guesses (\ref{e10}) permit 
higher statistical confidence (smaller error) of extracting {\it
both} path and phase information (with equal weights), than their
standard-case counterparts (\ref{e11a}).

In order to compare the information obtainable by the TIE-based and standard strategies, it is instructive to examine the
complementary quantities recently introduced \cite{Jakob} for entangled two-qubit systems:

(i) The {\em concurrence}, a measure of the two-qubit entanglement, becomes for the TIE wavefunctions (\ref{e4}) ($k_2/k_1=3$):
\begin{eqnarray}
C_{\rm TIE} & \equiv & | \langle (\psi_+ + \psi_-) | \sigma_y \otimes \sigma_y |
(\psi_+ + \psi_-)^* \rangle | \nonumber \\
&=& 2 \sqrt{p (1-p)} | \sin{\phi}|,
\label{conc}
\end{eqnarray}
where $\sigma_y$ is the appropriate Pauli matrix.
If we define the {\it analog} of path distinguishability $D_{S}$
\cite{ScullyNature} for TIE, \\$( P_{\ww})_{\rm TIE}\equiv(1-{\cal D})/2$, then, to $\delta \phi^2$ accuracy, $C_{\rm TIE}={\cal D}$ (cf. Eq.(\ref{e10}) for $p=1/2$). The peculiarity of TIE is
that ${\cal D}$ is {\it phase dependent}.

(ii) The {\em coherence}, alias the 
{\em generalized visibility} ${\cal V}$ defined in \cite{Jakob}, becomes for the TIE
states (\ref{e2}): ${\cal V}=\sqrt{(1-p)^2+p^2+2p(1-p)\cos{2\phi}}$ ($k_2/k_1=3$). {\em This measure
oscillates with $\phi$}. Hence, although, for any phase $\phi$
\begin{equation}
{\cal D}^2 + {\cal V}^2 = 1
\end{equation}
in accordance with the known complementarity relation
\cite{ScullyNature,Jakob}, ${\cal V}$ {\em does not
describe} the {\em amplitude} of the TIE nonsinusoidal interference
pattern (\ref{e5}), but rather the {\em purity} of the two-path state.
We may instead invoke the {\em customary} visibility
\cite{ScullyNature,RempeNature,Mei,Scully} $V=\max\,(P_+ -
P_-)=\max\,(P_{AB}) = 1$. This {\em global} (phase-independent)
measure would then yield, for phases such that ${\cal D} \simeq
1$: ${\cal D}^2+V^2 \simeq 2$, at odds with standard
complementarity! Yet this complementarity violation merely
demonstrates the inadequacy of the customary definitions for the TIE
{\em nonsinusoidal} interference pattern. 

We are therefore led to
conclude that the phase information stored in the TIE pattern
requires a new {\em operationally}-oriented measure. An adequate
measure is the {\em phase-sensitivity}, expressing the {\em
phase-derivative} of $(P_{AB})_{\downarrow}$ or
$(P_{AB})_{\uparrow}$ in (\ref{e7}) or $P_{\pm}$ in (\ref{e5}):

\begin{equation}
{\cal S}\equiv\left| \frac{2{\rm d} P_{\pm}}{k_{\rm max}{\rm d}L}
\right|=\frac{1}{3}|(p-1)\sin{\phi}-3p\sin{3\phi}|, \label{Sdef}
\end{equation}
where we have chosen $k_{\rm max}L=3k_1L=3\phi$. Because we are
interested in maximizing both ${\cal S}$ and ${\cal D}$, we
restrict $p$ to the region $p\in(1/2,1)$ and $\phi$ to the vicinity
of $\pi/2$. After eliminating $p$ from Eq.(\ref{conc}) we get the
following ellipse equation (for the choice $k_2=3k_1$)
\begin{equation}
\frac{\left[{\cal S}+\frac{\sin{\phi}+3\sin{3\phi}}{6}\right]^2}{\frac{|\sin{\phi}-3\sin{3\phi}|^2}{36}}+
\frac{{\cal D}^2}{|\sin{\phi}|^2}=1.\label{ellipsoid}
\end{equation}
Extending Eq.(\ref{ellipsoid}) to any integer ratio $k_2/k_1=N$ we get the {\em generalized complementarity relation} for TIE (at the respective optimal phase value)
\begin{equation}
\frac{\left( {\cal S} - \frac{N-1}{2N} \right)^2}{\frac{\left( N+1
\right)^2}{4N^2}}+{\cal D}^2 = 1.\label{generalellipse}
\end{equation}

The relation (\ref{generalellipse}) is our main result.
When $N=1$,
setting ${\cal D}=D_S$ and using the {\em same definition} of sensitivity as
for TIE (Eq.(\ref{Sdef})) we obtain, for the optimal phase $\phi\approx\pi/2$, 
the sensitivity ${\cal S}=V$ and recover the standard complementarity relation of Eq.(\ref{equality}).
\begin{figure}[h]
\centering
\includegraphics[width=0.6\linewidth]{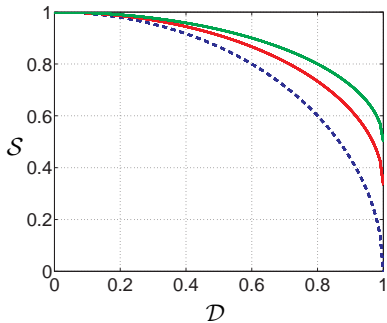}
\caption{Comparison of the ${\cal S}$, ${\cal D}$ dependence for standard case $N=1$, Eq.(\ref{equality}) (dashed-blue) and the TIE cases
Eq.(\ref{ellipse}) (red) and Eq.(\ref{ellipsehalf}) (green) at the optimal phases.}
\label{f2}
\end{figure}
The standard complementarity circle (\ref{equality}) ($N=1$) encloses a
smaller area in the ${\cal S}$-${\cal D}$ plane than the TIE complementarity
ellipse with $N>1$ (Fig.(\ref{f2})). This area difference
is a measure of the {\em additional information} on the paths stored in TIE
patterns compared to the standard case: higher ${\cal S}$ for the same ${\cal D}$
, or vice versa. As $N$ increases, so does the area
difference. 
The choice $N=3$ ($k_2=3k_1$) and
$\phi\approx\pi/2$, discussed in Eqs.(\ref{e5}-\ref{e10}), yields
\begin{equation}
\frac{({\cal S}-\frac{1}{3})^2}{\frac{4}{9}}+{\cal D}^2=1.\label{ellipse}
\end{equation}
For $N\gg 1$ we find the largest additional information
\begin{equation}
\frac{({\cal S}-\frac{1}{2})^2}{\frac{1}{4}}+{\cal D}^2=1.\label{ellipsehalf}
\end{equation}

TIE is realizable by forces changing the momentum depending on the
internal state as in Stern-Gerlach setups \cite{Scully}. TIE
interferometry may use e.g., molecules \cite{Buzek}. Here we discuss
the following realizations: (i) An optical TIE setup may involve an
{\em entirely birefringent} MZI. Polarization states $| 1 \rangle$
and $| 2 \rangle$ are then entangled throughout the MZI with
wavevectors $k_1$ and $k_2$, so that
$\phi_{A2}-\phi_{A1}$ or $\phi_{B2}-\phi_{B1}$ in (\ref{e7}) may
attain $\pi$ if $L_A, L_B$ exceed 0.1mm. (ii) An atomic realization
may involve the experimentally tested \cite{RempeNature} atomic MZI
based on Bragg scattering of a cold $^{85}$Rb atom from standing
light waves. We may envisage a cold $^{85}$Rb atom, moving
vertically along the $z$ -axis with momentum $\hbar k_z$, $\hbar k_x
= 0$, its internal state being the lowest hyperfine level, $F=2$. A
Ramsey RF field prepares the superposition between states $F=2$ and
$F=3$. Subsequently, as the atom moves through two travelling Bragg
gratings, each hyperfine state ``feels'' a different grating (by
tuning the field of each grating close to resonance with a different
electronic transition) such that the atoms in the states $F=2$ and
$F=3$ are Bragg-reflected to acquire, e.g., \emph{transverse} wavenumbers,
$k_1=k_x$ and $k_2=3k_x$, respectively. The atom is thereby prepared in the
TIE state (\ref{e1}), with $N=3$, which can then travel through the MZI.
At the output, one can project the internal states on a suitable
basis, by another Ramsey RF field.

The overlap of the wavepackets centered at $k_1$ and $k_2$ decreases as they propagate. This reduces our ability to distinguish the paths via the coherence between internal states $|1\rangle$ and $|2\rangle$, as per Eqs.(\ref{e4})-(\ref{e7}).
If either the $k_x$ or $3 k_x$ wavepacket has a length $w_{1(3)}$
not much larger than the MZI length $L$, so that their overlap is {\em
incomplete} at the output, the distinguishability will drop as
$\exp[-AL^2/w_{1(3)}^2]$, $A$ being a constant.

The crux of our new effects is that the TIE state (\ref{e1}) allows
us to perform {\em unconventional}
``quantum erasure'' \cite{Scully}, providing information on {\em
both} interfering paths at the expense of the internal states to
which they are entangled. Standard complementarity holds for
projection on {\it one} of the alternative paths $|A \rangle$ or $|B
\rangle$, hampering their superposition
\cite{Wooters,ScullyNature,RempeNature}. It needs to be generalized in the
present case (cf. (\ref{generalellipse})), where the 4D TIE state is projected onto an
internal-state (2D) basis. We may thus acquire {\it more
information}, by virtue of TIE, on any chosen 2D superposition of the $| A
\rangle$ and $| B \rangle$ path states. The resulting path and phase
information is {\it real} and {\it verifiable}. Such intra-particle
entanglement may become a new resource of quantum information or
interferometric measurements.


The support of EU (QUACS and SCALA), FRV\v{S}(2712/2005),
GA\v{C}R(202/05/0486) and ISF is acknowledged. We thank M. Arndt, S.
D\"{u}rr, B. G. Englert, G. Rempe, Y. Silberberg and A. Zeilinger
for useful discussions.

\hrulefill

\end{document}